\documentclass[12pt,preprint]{aastex}

\usepackage{natbib}
\usepackage{epsfig}
\newcommand{\fuse}{{\it FUSE\/}}
\newcommand{\euve}{{\it EUVE\/}}

\newcommand{\LA}{Lyman-$\alpha$}
\newcommand{\LB}{Lyman-$\beta$}
\newcommand{\LG}{Lyman-$\gamma$}
\newcommand{\kms}{km~s$^{-1}$}
\newcommand{\Htwop}{\mbox{H$\stackrel{+}{_2}$}}
\newcommand{\Htwo}{\mbox{H${_2}$}}

\begin{document}

\title{\fuse\ Observation of the Ultramassive White Dwarf PG~1658+441}

\author{Jean Dupuis, Pierre Chayer\altaffilmark{1}}
\affil{Department of Physics and Astronomy, The Johns Hopkins
University, 
\\ Baltimore, MD 21218}
\altaffiltext{1}{also with the Department of Physics and Astronomy,
University of Victoria, PO Box 3055, Station Csc, Victoria, BC V8W
3P6, Canada.}
\email{jdupuis@pha.jhu.edu, chayer@pha.jhu.edu}

\author{St\'ephane Vennes} 
\affil{Department of Mathematics, Australian National University,
\\ Canberra, Australia ACT0200}
\email{vennes@wintermute.anu.edu.au}

\and

\author{Nicole F. Allard\altaffilmark{2}, Guillaume H\'ebrard}
\affil{Institut d'Astrophysique de Paris, F-75014 Paris, France}
\altaffiltext{2}{also with Observatoire de Paris-Meudon, LERMA, F-92195 Meudon Principal Cedex, France.}
\email{allard@iap.fr, hebrard@iap.fr} 

\begin{abstract}

We present an analysis of the {\it Far Ultraviolet Spectroscopic
Explorer} ({\it FUSE}) spectrum of the ultramassive ($M$ = 1.31
M$_\odot$), magnetic ($B_s = 2.3$ MG) white dwarf PG~1658+441.  The
far ultraviolet (FUV) spectrum exhibits very broad Lyman lines and
quasi-molecular Lyman $\beta$ satellites, but weak Lyman $\gamma$
satellites may also be present.  PG~1658+441 is the hottest white
dwarf known to show these satellite features.  We fit the Lyman lines
with stellar models and obtain atmospheric parameters consistent with
a published analysis of the Balmer lines. By averaging results
obtained for the different \fuse\ segments, we determine $T_{\rm eff}
= 29,620\pm500$~K and $\log g = 9.31\pm0.07$. The models match the
data over large portions of the spectrum but discrepancies remain near
the satellite features.  Finally, no trace elements have been
identified in the FUV spectrum, and we provide abundance upper limits
for C, N, Si, and P.

\end{abstract}


\section{Introduction}

The DA white dwarf PG~1658+441 was
discovered in the Palomar-Green survey \citep{Green:1986}. 
\cite{Liebert:1983} announced the presence of a
magnetic field in this white dwarf and presented a detailed analysis
of the energy distribution consistent with a 30,000 K pure hydrogen
star. They measured a mean surface field
of $B_s= 2.3$ MG from the Zeeman splitting of the Balmer lines. The
magnetic field of PG~1658+441 was studied in detail by
\cite{Achilleos:1989} using offset magnetic dipole models. They measured a
field of $B_d = 3.0\pm0.5$ MG with a viewing angle between 40\degr\
and 50\degr. \cite{Schmidt:1992} reported the first surface gravity
measurement by fitting the Balmer line profiles with synthetic
profiles including Zeeman splitting. They reached the surprising
conclusion that PG~1658+441 has an abnormally high surface gravity of
$\log g = 9.36 \pm 0.07$ corresponding to a stellar mass of
$1.31\pm0.02$ $M_\odot$. The mass of PG~1658+441 appears much higher
than the average mass of field white dwarfs ($\approx
0.55-0.60M_\odot$). While searching for helium in ultramassive white
dwarfs, \cite{Dupuis:2002} analyzed the \euve\ spectrum of PG~1658+441
and found evidence for a small quantity of atmospheric contaminants
which was assumed to be helium.

The origin of PG~1658+441 and, more generally, of the population of
hot ultramassive white dwarfs discovered in recent EUV/soft X-ray
surveys is yet unknown.  These objects appear to form a secondary peak
in the white dwarf mass distribution near 1.25~$M_\odot$ and a
substantial fraction of these objects are also magnetic
\citep{Vennes:1999}.  As recently discussed by \cite{Dupuis:2002}, the
problem posed with the mass of these stars is that they exceed the
maximum mass of a C/O core ($\approx 1.1 M_\odot$) formed by the
evolution of intermediate mass stars
\citep{Weidemann:2000}. While it is not unconceivable to form white
dwarfs with O/Ne core with mass exceeding 1.1 M$_\odot$ through single
star evolution channels \citep{Nomoto:1984, Garcia-Berro:1997a},
\cite{Schmidt:1992} raised the issue that if PG1658+441 originates
from a star with a mass in excess of 6--8 M$_\odot$, it should be
relatively young and would be likely associated with a stellar
cluster. Because PG1658+441 is not an obvious member of any cluster,
\cite{Schmidt:1992} proposed instead that PG1658+441 could be
the product of the coalescence of a double-degenerate system. Such a
system is expected to rotate rapidly and show
photometric or magnetic field variability, but there is no
evidence yet that PG1658+441 is variable
\citep{Liebert:1983, Shtol':1997}. There is also the possibility that
PG1658+441 originates from an older Ap star, and that a peculiar
initial-mass to final-mass relation applies to the evolution of magnetic stars.

Why observe ultramassive white dwarfs with \fuse?  Atmospheric
parameters are usually obtained from a model atmosphere analysis of
Balmer line profiles. With recent advance in FUV astronomy,
spectroscopy of the hydrogen Lyman series has been made possible, and
an independent diagnostic of atmospheric properties of hot white
dwarfs is now available (see for example \citealt{Finley:1997b},
\citealt{Barstow:2001}, and \citealt{Barstow:2003}). The Lyman line
series also allows for a test of line broadening theory in the high
density conditions encountered in the atmosphere of ultramassive white
dwarfs. The high spectral resolution ($R = 17000-20000$) achieved with
\fuse\ allows for the detection of heavy element lines and helps
better characterize the atmospheric composition of ultramassive white
dwarfs. The detection of helium in the \euve\ spectrum of GD~50 by
\cite{Vennes:1996} indicates that trace elements may be present in
ultramassive white dwarf atmospheres. Although a study by
\cite{Dupuis:2002} shows that GD~50 is quite unique, the
case is still open due to the limited sensitivity of
\euve. \fuse\ observations of this class of objects are needed to
provide an independent
determination of their stellar parameters.

We present an analysis of the \fuse\ spectrum of the ultramassive
white dwarf PG 1658+441. In section 2, we describe the observations
and data reduction. In section 3, we present the model atmosphere
analysis and describe the abundance measurements of C, N, Si, and
P. We also analyze the Lyman line profiles which clearly show
the quasi-molecular satellite features of \Htwop\ in the wings of
Lyman-$\beta$ and Lyman-$\gamma$. This result is interesting in view
of the high surface gravity and relatively high effective temperature
of the star. We conclude by discussing the implications of our results
on the origin of ultramassive white dwarfs and we summarize our
findings.

\section{Observations and Data Reduction}

PG~1658+441 was observed as part of our \fuse\ Cycle 2 GI program
B122. The spacecraft and the instruments are described in
\cite{Moos:2000, Moos:2002} and an initial account of the calibration
is given in \cite{Shanow:2000}. \fuse\ provides spectra from 905 to
1187 \AA\ with a velocity resolution of 15--20 \kms\ in four
independent channels (SiC1, SiC2, LiF1, LiF2) each split in two
segments (A and B). The sequence of 16 exposures was initiated on 17
July 2001 at 13:42:45 (UT) for a total exposure time of 28483
seconds. The target was centered in the LWRS aperture and the
observations were made in TTAG mode. We have reprocessed the data with
CALFUSE version 2.2.2. We found the target to be out of the aperture
during 4 out of the 16 exposures in the LiF2 channel.

We have coadded the spectra of each channel from all exposures making
sure we selected only photometric exposures. The coaddition is made
using the program ``cf\_combine" provided with CALFUSE. Figures 1 and
2 show the combined spectra binned by 24 pixels for each segment of
the \fuse\ spectrometer. The Lyman lines series from the white dwarf
photosphere clearly dominate the spectrum.  A broad depression
centered on 1150 \AA\ was initially evident in the LiF1B spectrum, but
it was not observed in the LiF2A spectrum which covers a similar
wavelength range. This instrumental artifact, nicknamed the ``worm",
is thought to be due to blockage of the spectral image by a grid wire
above the surface of the micro-channel plate detector. We have derived
a ``worm" correction using the LiF2A segment as a reference
spectrum. We smoothed the ratio of the LiF2A to LiF1B spectra using
cubic splines and corrected the LiF1B spectrum by multiplying it by
this smoothed ratio.

\section{Analysis of the FUV spectrum of PG~1658+441}

In order to analyze the spectrum of PG~1658+441, we used
the model atmosphere code TLUSTY(v.~195) \citep{Hubeny:1988, Hubeny:1995}
and the spectral synthesis code SYNSPEC version~45 (SYNSPEC45) which
incorporates hydrogen Lyman line opacities \citep{Lemke:1997}, and
SYNSPEC version~48 (SYNSPEC48) which incorporates new
quasi-molecular satellite line opacities of \Htwo\
and \Htwop\ computed by N. Allard for \LA, \LB, and \LG. The
quasi-molecular opacity tables of \LB\ were specifically recomputed for a
temperature of 30,000 K, more appropriate for
PG~1658+441. 
The general method is described by \cite{Allard:1998} who computed theoretical line
profiles for \LB\ including quasi-molecular satellites due to \Htwop\
with a variable dipole. 

\subsection{Quasi-Molecular Satellites}

First, we computed a pure hydrogen model atmosphere in non-LTE
adopting published atmospheric parameters of PG~1658+441, $T_{\rm
eff}=30510\pm200$K and $\log g = 9.36\pm0.07$
\citep{Schmidt:1992}. Next, we calculated a synthetic spectrum using
SYNSPEC45 and covering the \fuse\ spectral range. This generic model
is normalized to match the absolute flux level in LiF1A and LiF1B. The
normalization corresponds to a $V$ magnitude of about 14.79 and agrees
with the same normalization based on {\it International Ultraviolet
Explorer} spectra
\citep{Dupuis:2002}. 

Figures 1 and 2 show a comparison between the model spectrum ({\it
solid lines}) and the \fuse\ spectra. The generic model seems in good
agreement with the observed line profiles, even though we have not yet
attempted a formal fit of the Lyman line profiles at this point (see
section 3.2). However, a closer inspection of Figures 1 and 2 shows
discrepancies in the \LB\ and \LG\ red wings.

Broad absorption troughs are apparent at 1058~\AA\ and 1076 \AA\ in
the red wing of \LB\ of the LiF1A, LiF2B, SiC1A, and SiC2B
spectra. The troughs are attributed to \Htwop\ quasi-molecular
satellite features previously detected in cooler white dwarfs by the {\it
Hopkins Ultraviolet Telescope} ({\it HUT})
\citep{Koester:1996b}, {\it ORFEUS} \citep{Koester:1998}, and \fuse\ 
\citep{Wolff:2001, Hebrard:2002}.  In addition, the features possibly
present on the red wing of \LG\ do correspond to quasi-molecular
features detected in the {\it HUT} spectra of WD1031$-$114 and
WD0644+375 by \cite{Koester:1998} and in the \fuse\ spectrum of
CD$-$38$^o$10980 by \cite{Wolff:2001} and included for the first time in
model comparisons presented by \cite{Hebrard:2003}.  PG~1658+441 is
the hottest white dwarf so far to show these FUV spectral features.

The validity of the Balmer and Lyman line profiles is questionable
because in high-gravity stars the Lyman line wings form at depths
where the electronic density is between $10^{17}-10^{18}$ cm$^{-3}$,
near the limit of Lemke's tables \citep{Lemke:1997}. Moreover, the
tables should not be applied directly to PG~1658+441 because they do
not include the satellite lines from quasi-molecules of \Htwop, but
the good agreement between the synthetic spectra and the observations
(Figs. 1 and 2) away from the quasi-molecular satellite features
suggests that they are a good approximation.

The emergence of quasi-molecular satellite lines of \Htwop\ in
PG~1658+441 is possibly due to its high surface gravity which implies
larger particle densities in the line forming region. Because
PG~1658+441 is hot, hydrogen should be more ionized but the high
surface gravity implies larger pressures (or larger electronic
densities) in the atmosphere which in turn favor
recombination. Because of this effect, we find the number density of
shydrogen atoms to be comparable with the one found in the atmospheres
of DA in which the quasi-molecular satellites lines have been
previously detected. These stars are typically cooler with $T_{\rm
eff}$ up to about 20,000K and less massive with $\log g$ close to
8. Because the quasi-molecular opacity is proportional to the proton
and \ion{H}{1} densities, we predict that the \Htwop\ quasi-molecular
satellite lines become apparent at higher effective temperatures in WD
with larger surface gravities.

Finally, adopting the same effective temperature and surface gravity
($T_{\rm eff} = 30,510$ K, $\log g = 9.36$), we computed a synthetic
spectrum using SYNSPEC48 which includes quasi-molecular satellite line
opacities.  Figures 1 and 2 show a comparison between the data and the
synthetic spectra which include the quasi-molecular satellite lines of
\LB\ and \LG\ ({\it dashed lines}). The predicted satellite lines
do not agree well with the data as the \Htwop\ satellites at 1058
\AA\ and 1076 \AA\ are somewhat too deep and shifted toward shorter
wavelengths. On the other hand, the predicted \LG\ satellites are too
weak. There is also a subtle kink predicted in the \LB\ profile near
1037 \AA\ which is possibly present in the data and better seen in
Figure 3 (see next section). This preliminary comparison is certainly
promising but we must investigate whether or not the agreement could
be improved by doing a formal fit of the \fuse\ spectrum.

\subsection{Atmospheric Parameters}

To determine the atmospheric parameters more accurately, we have
computed a fine grid of model atmospheres and synthetic spectra. The
spectra were computed with SYNSPEC48 which includes the
quasi-molecular opacities of \Htwop\ for \LB\ and \LG. The grid,
appropriate for determining the atmospheric parameters of PG~1658+441,
includes effective temperatures from 25,000 to 35,000K in steps of
1000K and surface gravities from 7.5 to 9.5 ($\log g$) in steps of 0.1
dex.

Each of the spectroscopic channels showing the Lyman lines are fitted
separately so that we obtain independent measurements of the stellar
parameters. We used a $\chi^2$--minimization technique to find the
best-fit parameters, $T_{\rm eff}$ and $\log g$.  We excluded regions
contaminated by airglow, interstellar absorption lines, as well as
detector edges where the spectra are of poor quality.  We neglected
Zeeman splitting which could produce shifts of the order of 1 \AA\ or
less in the core of \LB\ and higher Lyman lines. The approximation
does not affect the results because the line cores are already
excluded from the fit. We also binned the spectra by 20 pixels (which
corresponds to about 0.13\AA) to improve the signal-to-noise ratio.
For the fit of the \LB\ line, we restrict the fit to wavelengths
between 1005\AA\ and 1050\AA\ to exclude regions affected by the
quasi-molecular satellites of \LB\ and \LG. Including these regions
had the consequence of driving the fits toward spurious solutions. The
spectrum normalization (or scale factor) is also left as a free
parameter in the fit. An overall wavelength shift of $-0.4$ \AA\ was
applied to the model spectra to align optimally the models with the
data. Table 1 summarizes the results. The quoted error bars are
computed from the projection of the 3$-$$\sigma$ confidence contours
in the $T_{\rm eff}$--$\log g$ plane. A representative example of a
fit is shown on Figure 3 for the \LB\ line in the LiF1A segment which
shows how well the line is fitted up to where the quasi-molecular
satellites become important.

As expected, there is some scatter in the parameter values derived
from the different channels and segments which gives us an assessment
of the systematic errors. The measured parameter values are consistent
within the quoted error bars although the surface gravity measured
from SiC2B appears to be greater. This last measurement is less
reliable, as reflected by the larger error, because the \LB\ profile
from SiC2B is missing most of the blue wing (the wavelength coverage
ends at 1017 \AA).  Minor differences observed between different
channels are probably due to subtle flux calibration effects such as
light blocking by grid wires or sensitivity degradation which are not
perfectly accounted for in the data reduction.  Nonetheless, the
values obtained by taking the weighted mean of the six measurements
are in agreement with those of \cite{Schmidt:1992} although we seem to
favor slightly lower values of the gravity and effective temperature.

Our results confirm the analysis of \cite{Schmidt:1992} which was
possibly affected by Zeeman splitting observed in the Balmer line
cores. In general, they validate gravity measurements in low-field
white dwarfs which are based on the modeling of spectral line wings
computed at zero magnetic field. It is also interesting to note that
\cite{Barstow:2001} obtained a slightly lower value of the surface
gravity based on the FUV {\it HUT} spectrum of the ultramassive white
dwarf GD~50 compared to values based on optical spectra, but such
differences do not appear worst than for less massive DA stars in
their sample.  Note that \cite{Friedrich:1996} suggested that Stark
broadening due to an electric field parallel to the magnetic field may
be important in magnetic white dwarfs and that it may explain the
broadness of PG~1658+441 Balmer lines. However, their best fit to an optical
spectrum of PG~1658+441 is quite poor and may indicate that the effect
is overestimated.  We have not considered this effect.

\subsection{Metal Abundance Upper Limits}

A search for heavy elements in PG~1658+441 is motivated by the
detection of helium in the atmosphere of GD~50 \citep{Vennes:1996} and
by the possible detection of helium or other contaminants in the
extreme-ultraviolet spectrum of PG~1658+441 \citep{Dupuis:2002}.
These observations suggest that other ultramassive white dwarfs may
have some metals in their atmospheres. These metals may provide
additional clues required to better understand the origin of these
stars as they may be relics of the original composition of the
ultramassive white dwarfs. Various scenarios for producing
ultramassive white dwarfs predict hydrogen poor
atmospheres. Therefore, it is important to understand the nature of
ultramassive white dwarfs which appear to have hydrogen-rich
atmospheres.

Thus, we have searched for the presence of metals in the atmosphere of
PG~1658+441 but have found none of the usual suspects. Nevertheless,
we have used the spectra to set upper limits on elements likely to be
detected in \fuse\ spectra of white dwarfs, that is C, N, Si, and P.
We computed non-LTE synthetic spectra for each element selected over a
range of abundances varying from X/H = $10^{-9}$ to $10^{-5}$ in steps
of 0.25 dex. The strongest predicted lines (as shown on Fig. 4) are
used to compute a 3--$\sigma$ upper limit on the abundance. Note that
the upper limits are measured at the rest wavelengths of the selected
transitions, and that we corrected the wavelength scale for a
predicted gravitational redshift of 211 \kms. There is an additional
uncertainty related to our lack of knowledge of the radial velocity of
PG~1658+441. Figure 4 shows the spectra on a velocity scale with $v=0$
corresponding to the laboratory wavelength. The noise appears uniform
over a range of $\pm500$ \kms. The linear Zeeman effect is not taken
into account in the synthetic spectra, but it could shift the $\sigma$
components by $\pm360$ \kms\ (assuming a surface average field of 2.3
MG). Limiting the analysis to the un-shifted component, this effect
adds $\approx 0.5$ dex to the abundance upper limits.  However, the
spectra do not show any evidence for shifted components.  The results
are shown in Figure 4 and summarized in Table 2. The upper limits are
rather stringent and shows that PG~1658+441's atmosphere is metal poor
and is consistent with having a pure hydrogen atmosphere.

\section{Discussion and Summary}

The fact that no trace elements are detected in the hydrogen
atmosphere of PG~1658+441 is perhaps not so surprising. At such a high
surface gravity we do not expect radiative acceleration to support
detectable abundance of photospheric heavy elements
\citep{Chayer:1995}. 
On the other hand, the hydrogen-rich
composition of PG~1658+441 is quite interesting. 
The surface of white dwarfs emerging from the high-end
of the main-sequence progenitor mass range (8--10 M$_\odot$) 
is expected to be depleted of hydrogen and even probably of helium
\citep{Garcia-Berro:1997a}. It may suggest that there was a sufficient
amount of hydrogen left deeper in the envelope to form, over time, a hydrogen-dominated
atmosphere.  On the other hand, the progenitor of PG~1658+441 may have been
an early-A or late-B magnetic star which developed
a higher degenerate core mass because of the presence of a magnetic field.

Alternatively, the hydrogen-rich atmosphere of PG~1658+441 may suggest
that hydrogen was accreted from the interstellar medium as suggested
by \cite{Segretain:1997}. However, in the present case, accretion of
interstellar hydrogen may be inhibited by the presence of a magnetic
field. Using the formalism of \cite{Angel:1981} and assuming that
PG~1658+441 is accreting at the Eddington rate as argued by
\cite{Koester:1976}, we calculated that a mean equatorial field of
about 4136~G would be sufficient to prevent the accretion flow from
reaching the surface of the star (assuming a particle density of 1
cm$^{-3}$ and a velocity of 50 \kms). With a mean surface field of 2.3
MG \citep{Schmidt:1992}, it is unlikely that PG~1658+441 would accrete
significantly. Therefore, it is unlikely that the star entered the
white dwarf cooling sequence with an atmosphere devoid of hydrogen.

The detection of quasi-molecular satellite absorption in the
Lyman-$\beta$ and Lyman-$\gamma$ wings of PG~1658+441 shows that these
features can be observed even in hot white dwarfs and that they should
be included in the modeling of Lyman lines up to effective
temperatures higher than assumed in the past. Our analysis indicates
that the treatment of quasi-molecular satellite opacities could be
improved, particularly in the range of ionic densities encountered in
the atmosphere of ultramassive white dwarfs. Our current approach does
indeed have some limitations. For instance, the quasi-molecular
satellite profiles we use are based on an expansion in density
\citep{Allard:1994} valid for low densities of hydrogen atoms ($n \ll
10^{18}$ cm$^{-3}$). This approximation is appropriate for white
dwarfs with cooler and lower surface gravities but is less acceptable
for ultramassive white dwarfs. This might explain in part why the fits
shown in previous investigations \citep[for
examples]{Koester:1998,Wolff:2001,Hebrard:2003} appear somewhat
better. Another improvement would be to include a variable dipole
moment in the calculations of the satellites of Lyman-$\gamma$ as
suggested by \cite{Allard:2003}. This may enhance the strength of the
Lyman-$\gamma$ satellites and help improve the fit shown on Fig. 2. At
this stage, we have not included the quasi-molecular satellites of the
Lyman lines in our model atmosphere calculations because those
features are rather weak in comparison with the Stark wings at the
relatively high effective temperature of PG~1658+441.  In a future
investigation, we plan to include the quasi-molecular satellites in
the model atmosphere and interpolate in tables of profiles computed
for densities appropriate for ultramassive white dwarfs.

In summary, we have presented an analysis of the \fuse\ spectrum of the
ultramassive white dwarf PG~1658+441 and we find that:
\begin{itemize}
\item The \fuse\ spectra are well represented by a pure hydrogen
non-LTE model spectrum computed with the atmospheric parameters
determined from optical spectroscopy.
\item \LB\ satellite absorption has been detected at 1058~\AA\ and
1076~\AA\, making PG~1658+441 the hottest star in which these
features are detected.
\item Upper limits on the abundance of C, N, Si,
and P imply that PG~1658+441 has a highly pure hydrogen atmosphere.
\end{itemize}

\acknowledgements

This work is based on observations made with the NASA-CNES-CSA Far
Ultraviolet Spectroscopic Explorer. \fuse\ is operated for NASA by the
Johns Hopkins University under NASA contract NAS5-32985. This research
is funded by NASA grants NAG5-11570 and NAG5-11844.

\bibliographystyle{apj}
\bibliography{apj-jour,ms}

\begin{deluxetable}{cccc}
\tablecolumns{4}
\tablewidth{0pt}
\tablecaption{Atmospheric Parameters of PG~1658+441 from \fuse}
\tablehead{  
\colhead{Channel--Segment} & \colhead{$T_{\rm eff}$} & \colhead{$\log g$} &
\colhead{Wavelength Range}\\
 & \colhead{($10^{3}K$)} & \colhead{(c.g.s.)} & \colhead{(\AA)}}
\startdata
LiF1A & $29.4\pm0.20$ & $9.34\pm0.15$ & 988--1082 \\
LiF2B & $29.7\pm0.30$ & $9.28\pm0.20$ & 979--1074 \\
SiC1A & $29.4\pm0.30$ & $9.27\pm0.20$ & 1004--1090 \\
SiC1B & $30.0\pm0.60$ & $9.31\pm0.30$ & 905--992 \\
SiC2A & $30.9\pm0.40$ & $9.18\pm0.32$ & 917--1005 \\
SiC2B & $29.3\pm0.40$ & $9.46\pm0.34$ & 1017--1104\\ \hline
weighted mean     & 29.62 & 9.31 & \\
weighted standard deviation & 0.50 & 0.07 & \\
\enddata
\end{deluxetable}

\newpage

\begin{deluxetable}{cc}
\tablecolumns{2}
\tablewidth{0pt}
\tablecaption{Abundance Upper Limits for PG~1658+441 from \fuse}
\tablehead{  
\colhead{Ionic Species} & \colhead{$\log X/H$} \\}
\startdata
\ion{C}{3} & $\la -6.8$ \\
\ion{Si}{3} & $\la -7.7$ \\
\ion{P}{4}--\ion{P}{5}& $\la -6.0$ \\
\ion{N}{2} & $\la -6.5$\\
\enddata
\end{deluxetable}

\clearpage

\figcaption{\fuse\ spectra of PG~1658+441 in the LiF channels.
The spectra of PG~1658+441 in the LiF1 and LiF2 channels are dominated
by broad Lyman lines.  The data are compared to model spectra at
$T_{\rm eff} = 30,510$ K and $\log g = 9.36$ including quasi-molecular
lines for \LB\ and \LG\ ({\it dashed line}) and excluding the
quasi-molecular lines ({\it solid line}). The geocoronal
\ion{H}{1} emission is detected in the core of \LB\ and
\LG. The emission features in the LiF2A segment are spurious and
result from a local degradation of the micro-channel plate gain. Gaps
in wavelength coverage are apparent between segments 1A and 1B and
between segments 2A and 2B.}

\figcaption{\fuse\ spectra of PG~1658+441 in the SiC channels. Same as in Figure 1.}

\figcaption{Fit of the \LB\ line profile of PG~1658+441 in the LiF1A
segment. The upper part of the figure shows a comparison between the
LiF1A spectrum (in histogram) and the best fit model (solid line). The
fit is restricted to the region delimited by the two vertical dashed
lines. The \ion{C}{2} and \ion{O}{1} features are weak interstellar
lines and the absorption line near 1043\AA\ is a spurious feature due
to a detector dead spot. In the lower panel, we plot the residuals of
the fit which shows significant deviations at the locations of the
\Htwop\ quasi-molecular satellites of \LB\ and
\LG. The relatively narrow 1,2, and 3 sigma contours shown in the
lower part of the figure indicate that the solution is well constrained.}

\figcaption{Upper limits on metal abundances in \fuse\ spectrum of
PG~1658+441. The non-LTE synthetic spectra of C, N, Si, and P lines
are compared to the \fuse\ spectrum.  The spectra were computed with
abundances set to 3--$\sigma$ upper limits.  The synthetic and
observed spectra are plotted as a function of radial velocity centered
on the rest wavelength of the line transitions chosen to compute
abundance upper limits.  The velocity scale is corrected for the
predicted gravitational redshift of 211 \kms.}

\newpage

\begin{figure}[t]
  \figurenum{1}
  \plotone{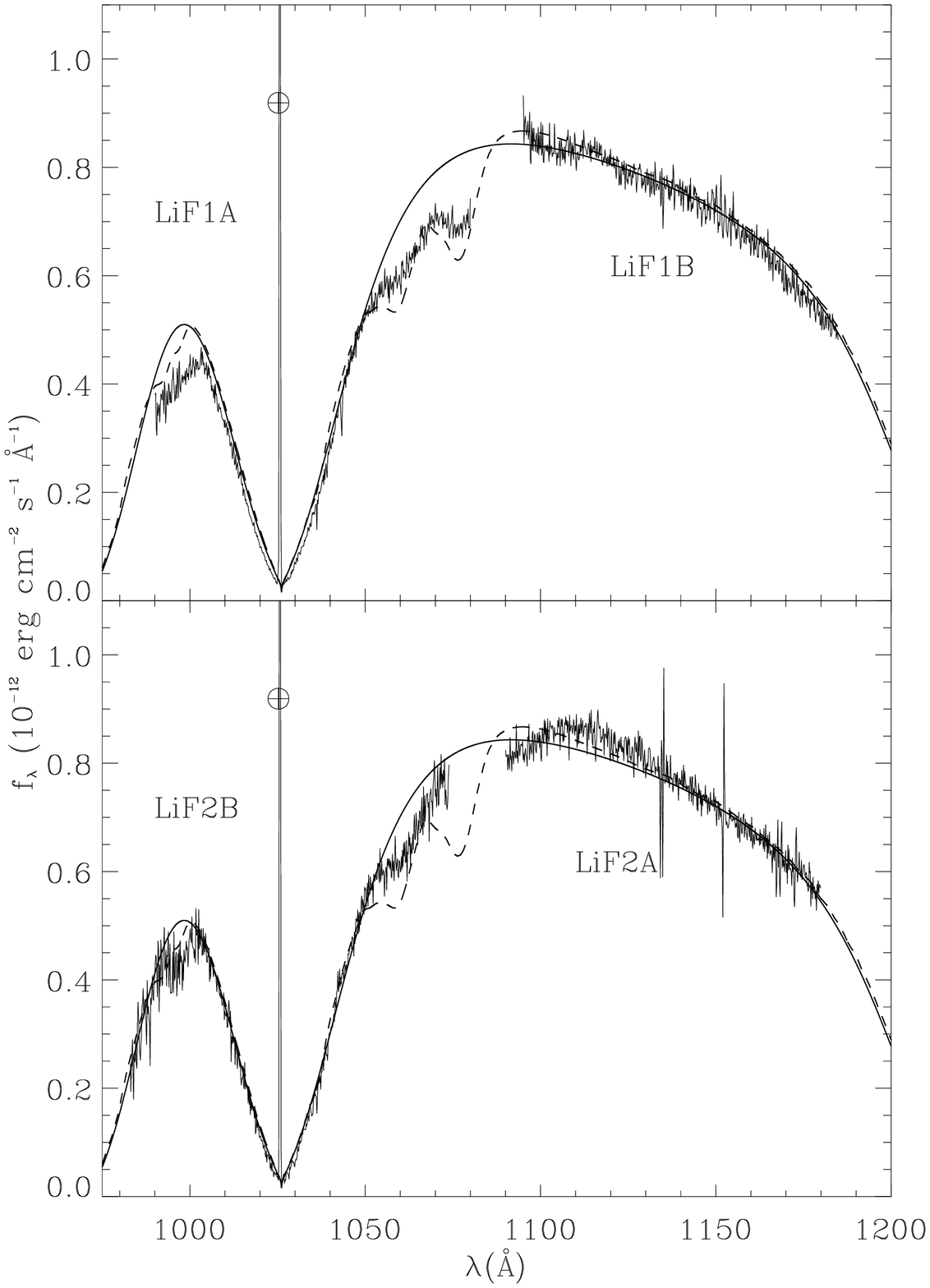}
  \caption{}
\end{figure}

\begin{figure}[t]
  \figurenum{2}
  \plotone{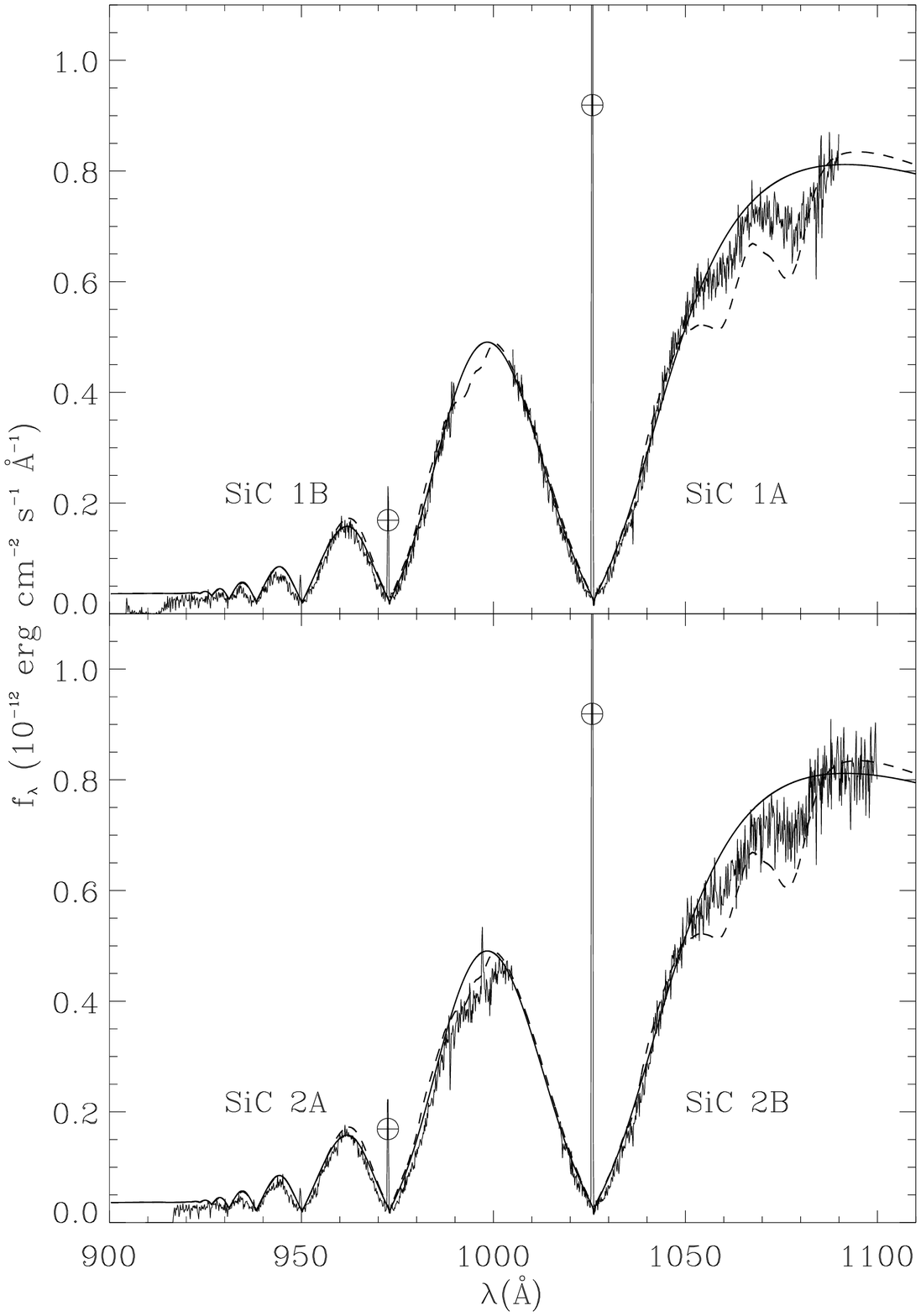}
  \caption{}
\end{figure}

\begin{figure}[t]
  \figurenum{3}  
  \plotone{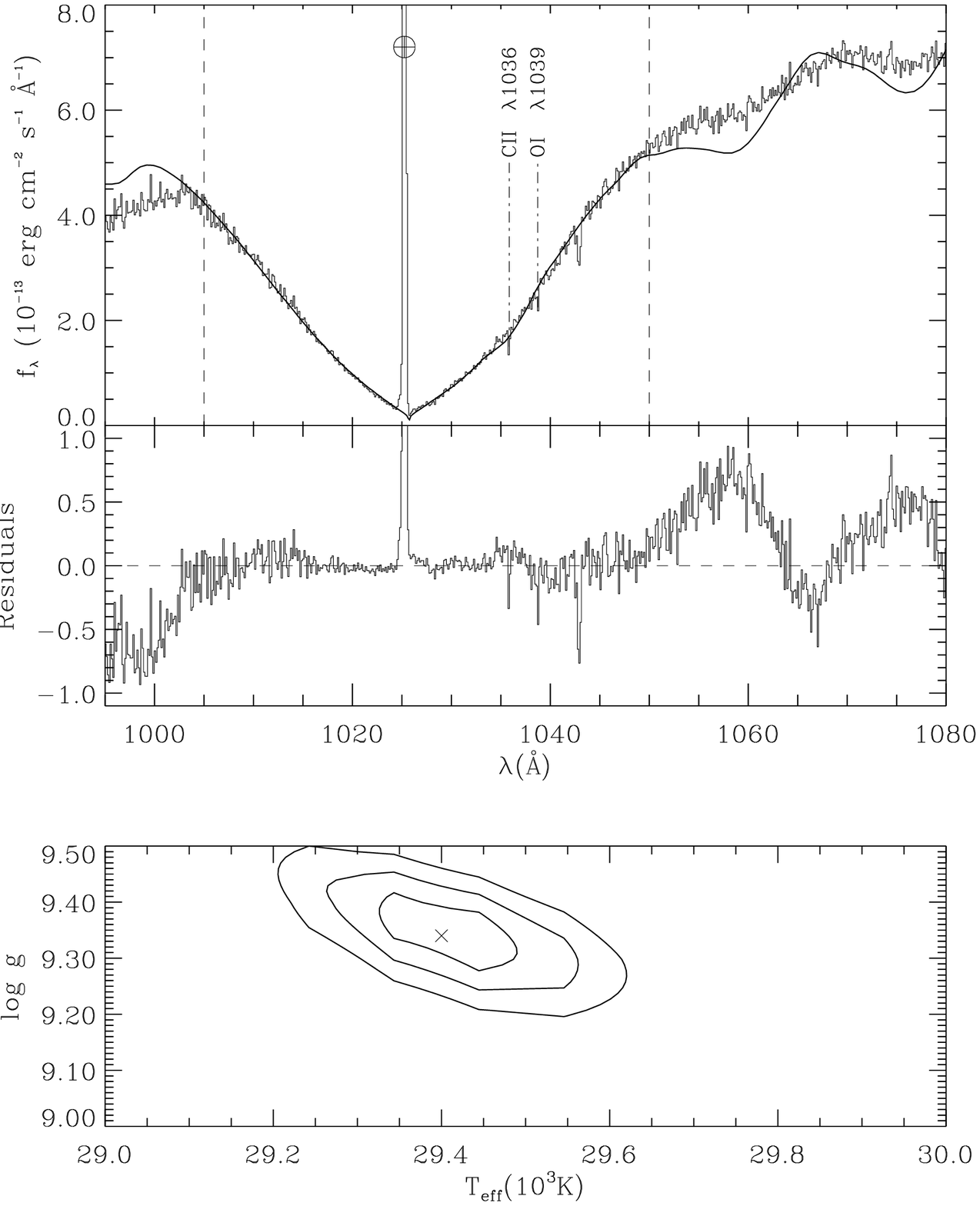}
  \caption{}
\end{figure}

\begin{figure}[t]
  \figurenum{4}
  \plotone{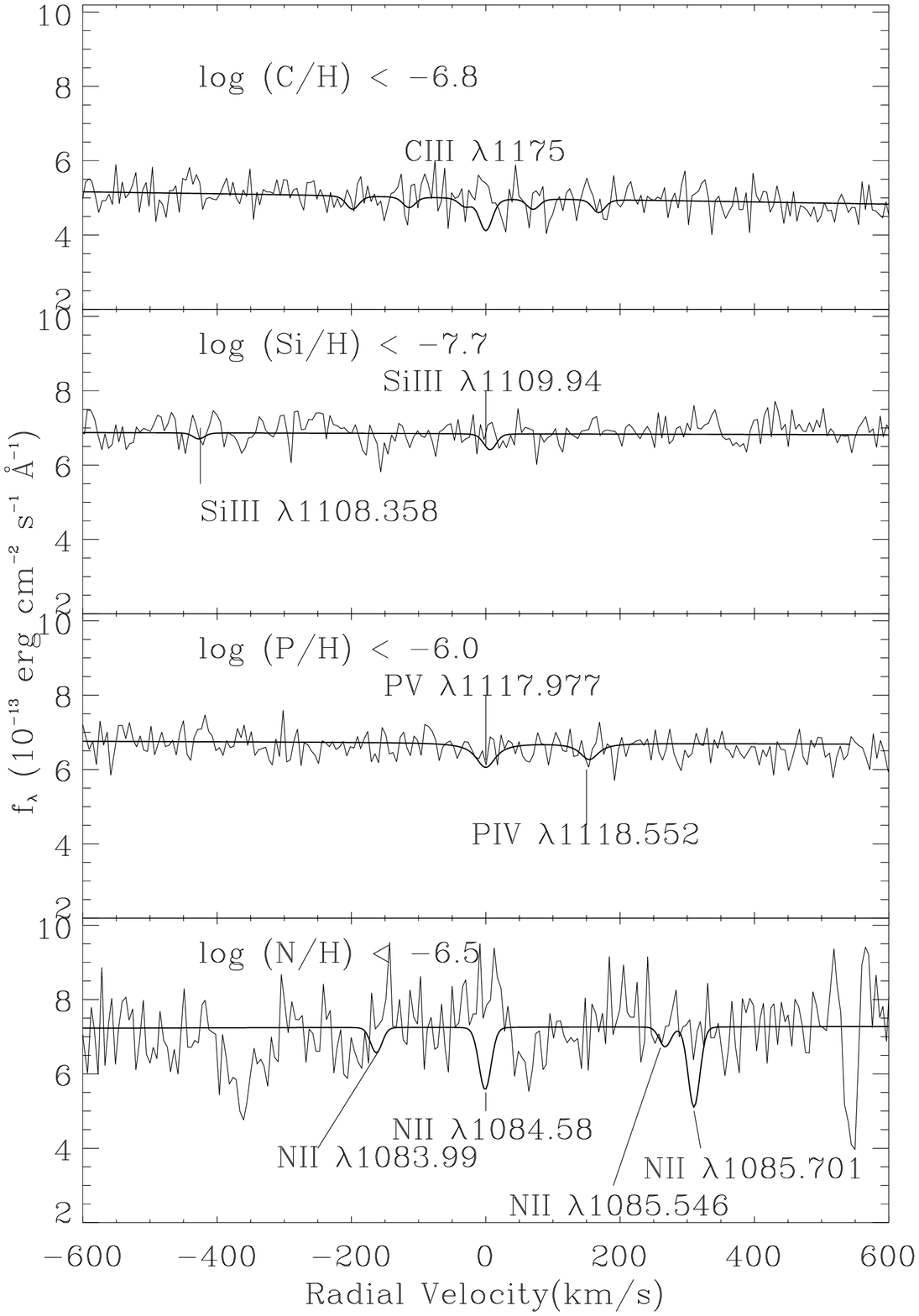}
  \caption{}
\end{figure}

\end{document}